\documentclass[prb,twocolumn,showpacs,floatfix]{revtex4-1}
\usepackage{graphicx}
\usepackage{dcolumn}
\usepackage{bm}

\begin{document}
\title{Hyperfine interaction in cobalt by high-resolution neutron spectroscopy}
\author{Tapan Chatterji$^{1*}$, Michaela Zamponi$^{2}$ and Joachim Wuttke$^2$}
\address{$^1$Institut Laue-Langevin, B.P. 156, 38042 Grenoble Cedex 9, France\\
$^2$Forschungszentrum J\"ulich GmbH, J\"ulich Centre for Neutron Science at MLZ, Lichtenbergstra\ss e 1, 85748 Garching, Germany\\
}
\email[Email of corresponding author: ]{chatterji@ill.fr}
\date{\today}

\begin{abstract}
We have investigated the ferromagnetic phase transition of elemental Co by high-resolution neutron backscattering spectroscopy. We monitored the splitting of the nuclear levels by the hyperfine field at the Co nucleus. The energy of this hyperfine splitting is identified as the order parameter of the ferromagnetic phase transition. By measuring the temperature dependence of the energy we determined the critical exponent $\beta = 0.350 \pm 0.002$ and the ferromagnetic Curie temperature of  $T_{\text{C}} = 1400$~K. The present result of the critical exponent agrees better with the predicted value (0.367) of the 3-dimensional Heisenberg model than that determined previously by NMR.
\end{abstract}
\maketitle

\section{Introduction}
It is well-known that the study of hyperfine interaction in magnetic solids yields valuable information about the electronic magnetic properties of solids.\cite{freeman65} There are several well-established techniques for studying hyperfine interaction in magnetic solids, viz.\ nuclear orientation and nuclear specific heats, the M\"ossbauer effect, nuclear magnetic resonance, angular correlation of $\gamma$-rays interaction of polarised neutrons with polarised nuclei, etc. Another less well-known method is to determine the hyperfine splitting of the nuclear levels directly by the spin-flip
scattering of neutrons.\cite{heidemann70} Unfortunately this element-specific technique is only suitable for a few isotopes with nuclear spins: They must have relatively large incoherent neutron scattering cross sections, whereas their absorption cross sections should not be prohibitively large. So far the study of hyperfine interaction by high-resolution neutron spectroscopy is limited to Nd \cite{chatterji00,chatterji02,chatterji04,chatterji04a,chatterji08,chatterji08a,chatterji09f} , Co\cite{heidemann75a,heidemann75b,chatterji09c,chatterji09d,chatterji10,chatterji10a}, V \cite{heidemann70,heidemann72,heidemann73,heidemann76,heidemann77} and to some extent also Ho \cite{chatterji13,chatterji13a,ehlers09}.
    
In the past, we studied low energy nuclear spin excitations in Nd and several Nd-based compounds,\cite{chatterji00,chatterji02,chatterji04,chatterji04a,chatterji08,chatterji08a,chatterji09f} and found that the ordered electronic magnetic moment of Nd$^{3+} $ ions are linearly proportional to the energy of excitations or the hyperfine splitting of the Nd nuclear levels. Out of seven naturally occuring isotopes $^{143}$Nd and $^{145}$Nd with natural abundances 12.18\% and 8.29\% respectively, have nuclear spin $I =7/2$ and their spin incoherent neutron scattering cross sections\cite{sears99} are rather large ($55 \pm 7$ and $5 \pm 5$ barn for $^{143}$Nd and $^{145}$Nd, respectively) making high-resolution neutron scattering investigation feasible. Another interesting candidate for such study is cobalt that has 100\% natural abundance of the isotope $^{59}$Co which has a large nuclear spin $I = 7/2$ and a relatively large incoherent scattering cross section\cite{sears99} of $4.8 \pm 0.3$ barn. Therefore Co and Co-based compounds are suitable for neutron spectroscopy. In fact Heidemann et al.\cite{heidemann75a,heidemann75b} studied hyperfine splitting in ferromagnetic Co and Co-P amorphous alloys and also Co-based intermetallic componds LaCo$_{13}$, LaCo$_5$, YCo$_5$ and ThCo$_5$. We reported the results of our recent studies on a number of Co compounds,\cite{chatterji09c,chatterji09d,chatterji10,chatterji10a} viz. CoO, CoF$_2$, Co$_2$SiO$_4$ and CoV$_2$O$_6$.

The hyperfine interaction study by high resolution neutron spectroscopy can be sometimes very useful because of its sensitivity to only a few magnetic atoms with nuclear spins. This element-specific methods is particularly  suitable for the study of two-sublattice magnetic structures and phase transitions. If only one of these two magnetic atoms yield good hyperfine signal then we can make clear statement of the magnetic ordering of that particular magnetic atom selectively.   

Hyperfine interaction in the elemental 3d ferromagnets Co, Fe, Ni is of particular interest. The seminal NMR study of these metals was done by Shaham et al.\cite{shaham80} Heidemann\cite{heidemann75a} studied the hyperfine interaction in metallic Co by high-resolution neutron spectroscopy only at room temperature. During the present investigation we studied the temperature dependence of the low energy nuclear spin excitations in ferromagnetic Co in a wide temperature range 3.5--1421~K. The ferromagnetic Curie temperature of cobalt is\cite{shaham80} as high as $T_{\text{C}}\approx 1395$ K and therefore the study of hyperfine interaction covering temperatures up to $T_{\text{C}}$ is technically rather demanding.

\begin{figure}
\resizebox{0.5\textwidth}{!}{\includegraphics{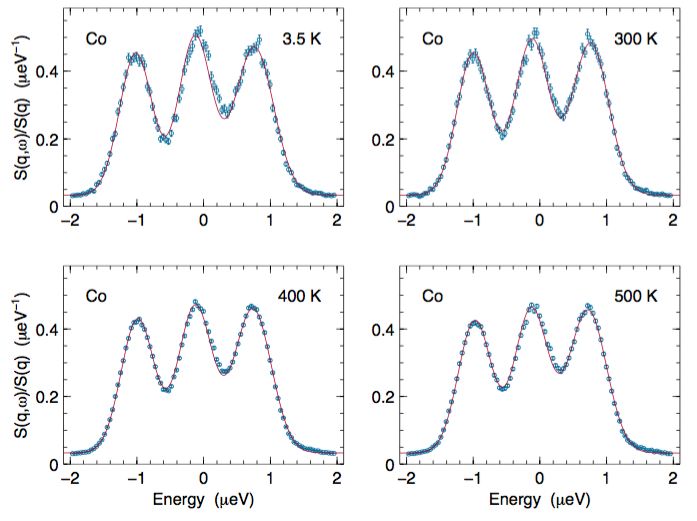}}
\caption {(Color online) Typical inelastic spectrum from cobalt at several temperatures in the low-$T$ range (Experiment~2). Solid lines are fits as described in the text.}
\label{Co-LTrange-spectra}
\end{figure}

\begin{figure}
\resizebox{0.5\textwidth}{!}{\includegraphics{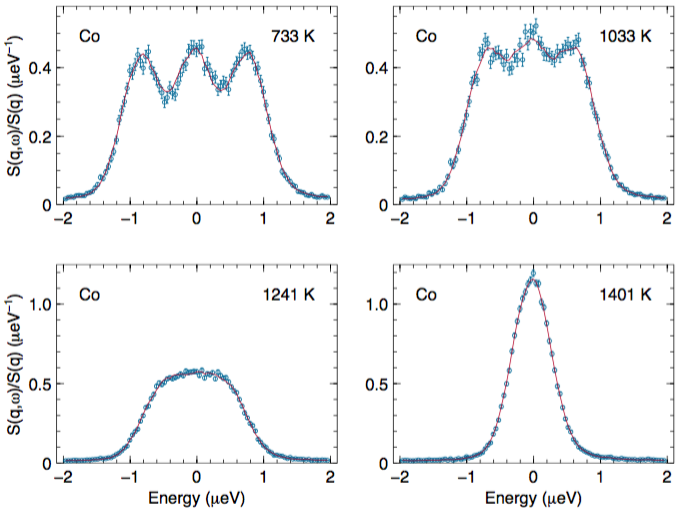}}
\caption {(Color online) Typical inelastic spectrum from cobalt at several temperatures
in the high-$T$ range (Experiment~3).}
\label{CoFig1}
\end{figure}

\begin{figure}
\resizebox{0.5\textwidth}{!}{\includegraphics{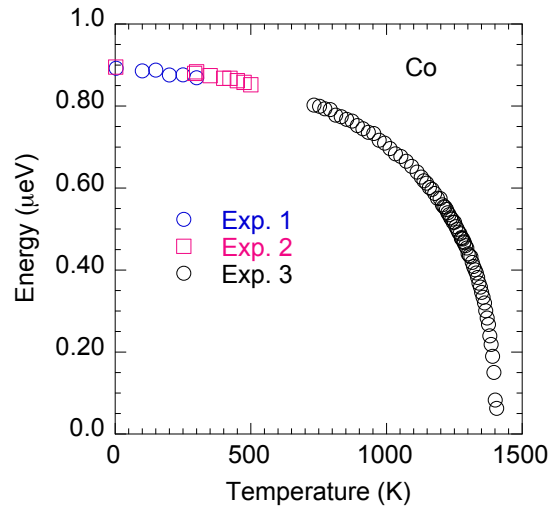}}
\caption {(Color on line) Temperature variation of the energy of nuclear spin excitations in cobalt in the whole measured temperature range. }
\label{Co-ETdep}
\end{figure}

\begin{figure}
\resizebox{0.5\textwidth}{!}{\includegraphics{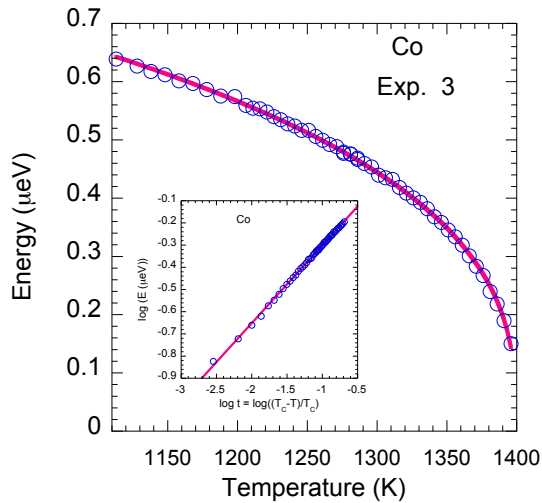}}
\caption {(Color on line) Temperature variation of the energy of nuclear spin excitations in cobalt in the high temperature range. The power-law fit (continuous red line) was done only with the data in the temperature range 1113 - 1398 K range. (Inset) Log - log plot of the data and the fitted straight line. }
\label{CoFig2}
\end{figure}

\begin{figure}
\resizebox{0.5\textwidth}{!}{\includegraphics{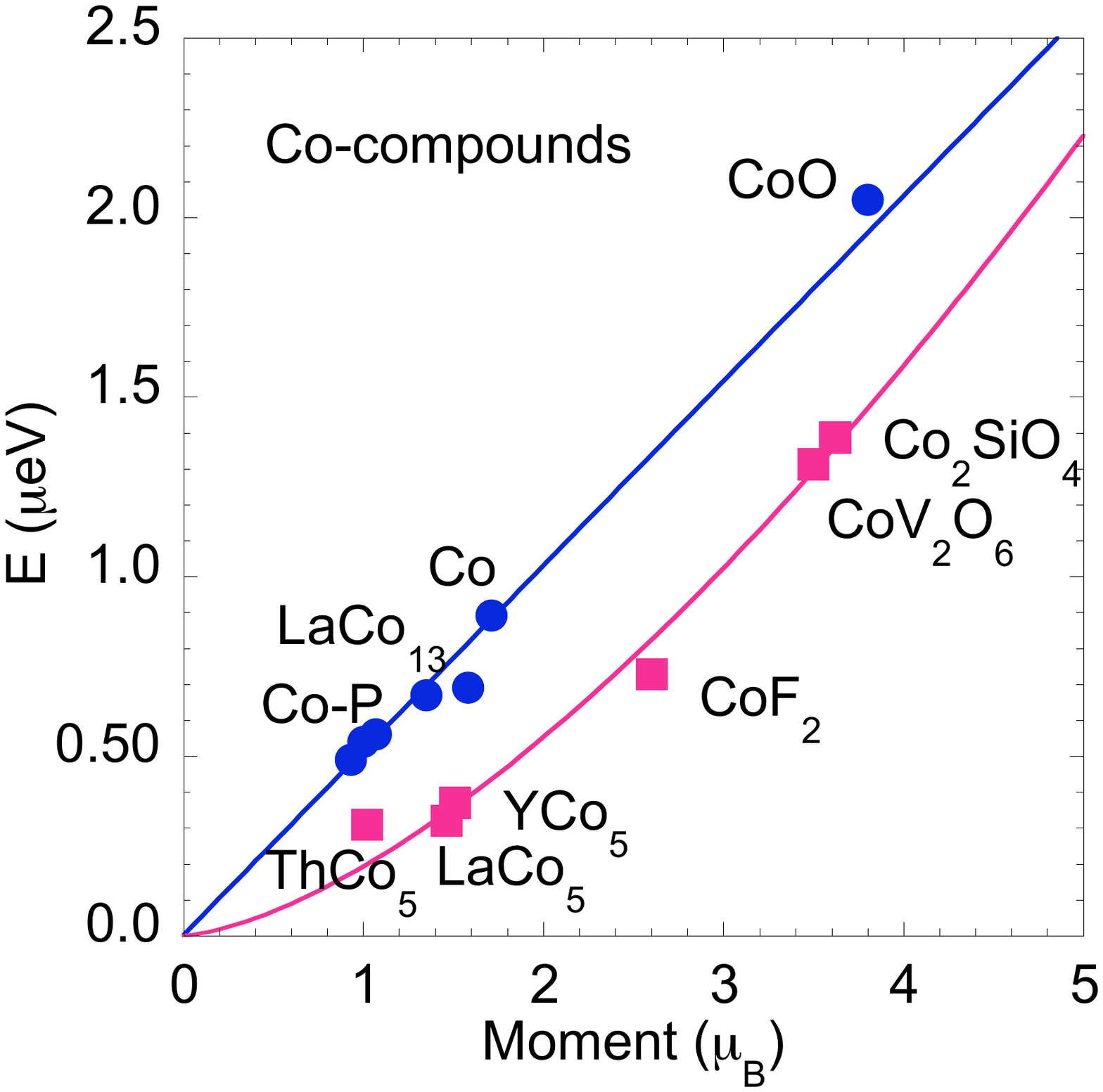}}
\caption {(Color online) Energy of hyperfine excitations in several Co-compounds as a function of the ordered magnetic moment.}
\label{CoFig3}
\end{figure}

\section{Experimental methods}
We performed inelastic neutron scattering experiments on cobalt by using the high-resolution neutron backscattering  spectrometer SPHERES\cite{spheres} of the J\"ulich Centre for Neutron Science (JCNS) located at the Heinz Maier-Leibnitz Zentrum in Garching, Germany. The instrument operates with a fixed final neutron wavelength of $\lambda= 6.271$~{\AA}.

Two samples and two sample environments were used to measure hyperfine spectra over a wide range of temperatures. Experiment~1 and 2 were carried out on a thin plate of polycrystralline Co ($25 \times 50 \times 0.5$ mm$^3$ ) placed inside a top-loading cryostat, covering the temperature range 3--500~K. In Experiment~3, a plate-shaped polycrystalline Co sample with size $25 \times 50 \times 0.5$ mm$^3$ was placed inside a furnace that could generate temperatures up to about 1800~K. In Experiments 1 and 2 measurements were done at a few selected temperatures. In Experiment 3 slow continuous ramps were chosen and the acquired time-resolved spectra later on binned together to obtain the different temperature points.



The sample plates were mounted in 45$^\circ$ reflection geometry. 
Since small-angle detectors have reduced energy resolution,\cite{spheres2}
they were discarded from the present analysis.
The large-angle detectors, covering scattering angles $2\theta$ from 35 to 134$^\circ$ 
all yielded spectra of comparable quality.
As expected for incoherent scattering,
besides a slight intensity variation, no systematic dependence on~$2\theta$ was observed.
Therefore, data from the 4 best 
detectors were all summed up
so that we obtained just one spectrum per temperature interval.

In quasielastic scattering experiments,
it is customary to determine the instrumental resolution
from a low-temperature spectrum where all relaxational dynamics is frozen so
that scattering from the sample is purely elastic.
In hyperfine studies, it is just the opposite:
The hyperfine splitting persists down to the lowest temperatures.
However, it vanishes at \emph{high} temperatures, above~$T_{\text{C}}$.
Therefore, to determine the resolution function we measured a spectrum far above~$T_{\text{C}}$, for a temperature range 1411 - 1421~K,  at which only elastic scattering was observed.
We obtained a full width at half maximum (FWHM) of 0.67~$\mu$eV,
in good agreement with the standard performance of SPHERES.\cite{spheres}

\section{Results}
Figs.\ \ref{Co-LTrange-spectra} and~\ref{CoFig1} show typical spectra from Co at different temperatures from 3.5 to 1401~K. The hyperfine splitting appears as a pair of inelastic peaks plus a central elastic peak. The three peaks should be of equal intensities. However,  the spectra we observed are slightly asymmetric.
The detailed balance factor $\exp(E/k_{\text{B}}T)$ is negligible throughout our temperature range.
The asymmetry is solely due to instrumental imperfections
that have to do with the intrinsic asymmetry of energy selection by backscattering.\cite{spheres2}
Also, the normalization to the incident neutron flux is only approximate.

For experiment 3, solid lines show fits with an elastic delta line and a pair of inelastic delta lines,
all convoluted with the instrumental resolution determined at high temperature plus a constant for the background.
The fit enforces the inelastic lines to have the same central energy $E$ with one amplitude for all three peaks. The  spectra obtained in the Experiment 1 and 2 had to be fitted with different intensities.

At $T = 3.5$~K, we obtain a hyperfine splitting of $E = (0.892 \pm 0.004) \mu$eV. This agrees well with the extrapolated low-temperature value of NMR frequencies determined by Shaham et al.\cite{shaham80} The inelastic peaks move towards the central elastic peak with increasing temperature and finally merge with it at the ferromagnetic phase transition temperature $T_C = 1400$~K.
Consequently at temperatures closer to $T_C$ the three-peak structure is not clearly seen, but they look like  broad peaks. However three peaks could still be fitted to the broad peak with some constraints as already discussed before.
Figs.\ \ref{Co-ETdep} show the temperature variation of the hyperfine splitting $E(T)$ measured in the whole temperature range. It decreases continuously as the temperature is increased.
 The ferromagnetic Co is known to undergo a structural phase transition at about 707 K from the low-temperature hexagonal to the high-temperature face centered cubic phase. There is a detailed neutron diffraction study of this structural phase transition \cite{frey79}.  However the magnetic properties hardly change at this transition and therefore the transition has not affected our measurements. The magnetization\cite{crangle71} and the NMR frequency\cite{shaham80} vary smoothly with temperature without showing any abrupt change of slope at this transition temperature and so does the energy of hyperfine splitting measured here.
On approaching~$T_{\text{C}}$, the inelastic lines merge with the central, resolution-broadened elastic peak. 

Fig. \ref{CoFig2} shows the temperature variation of $E(T)$ in the high temperature range. From our previous studies \cite{chatterji09d,chatterji10} we know that the energy of hyperfine field splitting $E(T)$ can be identified as the order parameter of the ferromagnetic phase transition. So we have fitted the temperature variation of the splitting energy by a power law given by 
\begin{equation}
      E(T) = A\left[(T_C -T)/T_C \right]^\beta
 \end{equation}
where $\beta$ is the critical exponent. The inset of Fig. \ref{CoFig2} shows a log-log plot of $E(T)$ vs. $\left[(T_C -T)/T_C \right] $ giving a straight line.  For this plot we used only the data in the range 1113-1398 K which are closer to $T_C$. From the least-squares power-law fit using the data from the above mentioned range, we also determined the ferromagnetic Curie temperature of Co to be $T_{\text{C}} = 1400$ K, which was 5 K higher compared to the reported\cite{shaham80} literature value $T_{\text{C}} = 1395$ K. The fit gave the critical exponent as $\beta = 0.350 \pm 0.002$ and $T_{\text{C}} = 1400.0 \pm 0.3$~K. The given error in $T_C$ is just the result from the least-squares fit. We know however that the real systematic error in temperature can be as high as $\pm 5$~K. The critical exponent is close to the predicted value $\beta =0.367 \pm 0.002$ of the critical exponent for a three dimensional Heisenberg system\cite{collins89,nonomura16}. 
There is a slightly different value  ($0.38 \pm 0.03$) given in other reference\cite{lindgard78} but with a much larger error.

\section{Discussion}
We have determined the order parameter of the ferromagnetic phase transition in elemental Co by high-resolution neutron backscattering spectroscopy. We have shown that this technique is capable of investigating such magnetic phase transition with a very high accuracy comparable to any other well-known conventional technique such as NMR and other competing techniques.The NMR data\cite{shaham80} for Co gave a much lower value of critical exponent $\beta = 0.309 \pm 0.012$, which do not agree at all with the expected three dimensional Heisenberg value. Neither this value for Co agrees with the critical exponent values determined by the same authors for Ni and Fe.
 We have also studied a whole series of other Co compounds by the same technique. In the following we will attempt to correlate the hyperfine splitting energy of Co with its electronic magnetic moment determined by mostly neutron diffraction.

 \begin{table}[ht]
\caption{Ordered electronic moment of Co and the energy of Co nuclear spin excitations in Co and other Co compounds. }
\label{table1}
\begin{center}
\begin{tabular}{llll} \hline \hline
Compound & Moment ($\mu_B$)&$\Delta E$~($\mu$eV) & Reference\\ \hline
CoV$_2$O$_6$& 3.5(1)& 1.379(6)&[15]\\
Co$_2$SiO$_4$& 3.61(3)& 1.387(6)&[14]\\
CoF$_2$ & 2.60(4)&0.728(8)&[13]\\
CoO & 3.80(6)&2.05(1)&[12]\\
Co & 1.71 & 0.892(4)&[10, present work]\\
Co$_{0.873}$P$_{0.127}$ & 1.35&0.67&[10]\\
Co$_{0.837}$P$_{0.161}$ &1.0& 0.54&[10]\\
Co$_{0.827}$P$_{0.173}$& 1.07&0.56&[10] \\
Co$_{0.82}$P$_{0.18}$ & 0.93&0.49 &[10]\\
LaCo$_{13}$&1.58&0.69&[11]\\
LaCo$_{5}$&1.46&0.32&[11]\\
YCo$_{5}$&1.51&0.37&[11]\\
ThCo$_{5}$&1.02&0.31&[11]\\\hline
\end{tabular}
\end{center}
\end{table}

Table I gives the ordered electronic moment of the Co and the energy of Co nuclear spin excitations of metallic Co determined during the present investigations along with the similar data obtained by Heidemann\cite{heidemann75a} in Co and Co-P amorphous alloys and by Chatterji et al.\cite{chatterji09c,chatterji09d,chatterji10,chatterji10a} in CoO, CoF$_2$, Co$_2$SiO$_4$ and CoV$_2$O$_6$ and by Heidemann et al.\cite{heidemann75b} in LaCo$_{13}$, LaCo$_5$, YCo$_5$, and ThCo$_5$. Fig.~\ref{CoFig3} shows a plot of energy of inelastic peaks observed in Co and Co compounds vs. the corresponding saturated
electronic magnetic moment of Co in these compounds. The data corresponding to Co, Co-P amorphous alloys, CoO and LaCo$_{13}$ lie approximately on a straight line
showing that energy of inelastic peak or the hyperfine splitting of the nuclear level is approximately
proportional to the electronic magnetic moment. The slope of the linear fit $E = a\mu$ ($\mu$ = magnetic moment) of all
data for Co, Co-P amorphous alloys, CoO and LaCo$_{13}$ gives a value of $a=0.51 \pm 0.01$~$\mu$eV$/\mu_B$. The data for all other compounds,  LaCo$_5$, YCo$_5$, and ThCo$_5$, CoF$_2$, Co$_2$SiO$_4$ and CoV$_2$O$_6$ do not fit at all with the straight line. The data for these \emph{anomalous compounds} can be fitted by by a power law $E=a\mu^n$ with $a = 0.19 \pm 0.03$ and $n= 1.5 \pm 0.1\approx 3/2$. The deviation from the linear behaviour for these \emph{anomalous compounds} is likely related with the existence of different orbital moments of Co ions in different compounds.

For an atomic nucleus with magnetic moment, there is a magnetic hyperfine interaction between the nucleus and the electron in addition to the usual Columbic interaction between them. The hyperfine interaction is caused by the magnetic field (hyperfine field $B_{\text{hf}}$) produced by the spin and the orbital moments on the nucleus. The hyperfine fields Co in hcp, fcc and bcc structures have been calculated by Guo and Ebert\cite{guo96}. The calculations were based on the first-principles relativistic, spin-polarised density functional theory.\cite{macdonald79} In the lowest order, the relativistic hyperfine interaction operator reduces to three terms, viz.\ the Fermi-contact, magnetic dipole, and orbital terms. The authors have calculated the magnetic moments and hyperfine fields separately for all these contributions. The calculated spin, orbital and dipole moments for the hcp form are 1.596, 0.077 and $-0.006$~$\mu_B$, respectively. The calculated total hyperfine field is $-22.52$~T of which core electrons contribute $-19.05$~T and the valence electrons contribute $-3.45$~T. The valence electron contribution $-3.45$~T consists of $-8.52$~T from s electrons and $+5.05$~T from non-s electrons. The non-s part 5.05~T can be identified to the orbital part of the hyperfine field. Unfortunately similar first-principle calculations are not yet available for the compounds listed in Table 1. However the experimental results presented in Table 1 may hopefully induce some first-principle calculations on these Co-compounds. Without such calculations it is not possible to understand the unusual behaviour of the hyperfine interaction shown in Fig.~\ref{CoFig3}.

\section{Conclusion}
In conclusion, we have measured hyperfine interaction in Co as a function of temperature covering an wide temperature range from 3.5--1421 K. We have shown that the less-known high-resolution neutron spectroscopic technique is capable of yielding good quality data comparable to that obtained by the well-known techniques like NMR. The present data of the hyperfine splitting in Co together with the data on other Co compounds obtained by high-resolution neutron spectroscopy show that there exist unresolved problems about the hyperfine interaction of Co compounds most likely due to the presence of orbital moments.

\section*{Acknowledgements}
The present work is based upon experiments performed on the SPHERES instrument operated by JCNS at the Heinz Maier-Leibnitz Zentrum (MLZ), Garching, Germany. The author TC gratefully acknowledges the financial support provided by JCNS to perform experiments at MLZ, Garching, Germany. We thank Simon Eder, Stefan Lindemaier and Milan Antic of the MLZ sample environment group for the design and construction of the high temperature furnace used in the experiment. We also thank Harold Schneider for his help during the measurements.

\end{document}